\documentclass{PoS}

\newcommand{\mf}[1]{\mathbf{#1}}
\newcommand{\ov}[1]{\overline{#1}}
\newcommand{\be}{\begin{equation}}
\newcommand{\ee}{\end{equation}}
\newcommand{\bey}{\begin{eqnarray}}
\newcommand{\eey}{\end{eqnarray}}

\usepackage[intlimits]{amsmath}
\usepackage{amssymb}
\usepackage{bbm}
\usepackage{mathrsfs}
\usepackage{dsfont}

\title{Excited state spectroscopy in the lattice Gross-Neveu model}

\ShortTitle{Excited state spectroscopy in the lattice Gross-Neveu model}

\author{\speaker{Julia Danzer}, Christof Gattringer \\ \\
        Karl-Franzens Universit\"at Graz, Austria, Institut f\"ur Physik\\ \\
        E-mail: \email{julia.danzer@edu.uni-graz.at,\\
\hskip12mm christof.gattringer@uni-graz.at}
}

\abstract{We present preliminary results of an excited state spectroscopy
  calculation in the 2-d lattice Gross-Neveu model. We address the
  construction of suitable interpolators for the variational method and their
  overlap with excitations. We comment on the role of the eigenvectors as a
  tool for matching scattering states on lattices with different volumes.} 

\FullConference{The XXV International Symposium on Lattice Field Theory\\
		 July 30 - August 4 2007\\
		 Regensburg, Germany}

\begin{document}

\section{Introduction}

\noindent
Lattice QCD is by now a mature field which produces reliable quantitative
results from ab-initio calculations. With refined techniques the range of
quantities that can be computed on the lattice is increasing continuously.  

Observables that recently have started to see a lot of attention in 
the lattice community are masses of excited states of light hadrons (see
e.g.~\cite{hadrons}). Excitations, however, are quantities that are
notoriously difficult on the lattice. The reason is that the energies $W_n$  
of excited states appear only as sub-leading exponentials in the spectral
decomposition of Euclidean 2-point functions,
\begin{equation}
C_{ij}(t) \; = \; \langle O_i(t) O_j(0)^\dagger \rangle \; = \; 
\sum_n \langle 0 | \widehat{O}_i | n \rangle 
\langle n | \widehat{O}^\dagger_j | 0 \rangle \, e^{-t \,W_n} \; .
\label{twopoint}
\end{equation}
Here $\widehat{O}_i, \widehat{O}_j$ are operators with the quantum numbers of
the state one is interested in. The sum runs over all physical states $| n
\rangle$ with these quantum numbers, and $W_n$ are the corresponding
energies ($W_1 < W_2 < W_3 \, ...\, )$. 
It is obvious, that the exponential decay of the 2-point function    
is dominated by the energy of the ground state $W_1$, while the energies of
the excited states $W_2, W_3 ...\,$ appear in sub-leading exponentials and thus
their contribution to the 2-point function is exponentially suppressed
relative to the ground state. 

Although other proposals for the extraction of the excites state energies
exist \cite{otherextract}, the most commonly applied technique is the variational
method \cite{variation1,variation2}. Here one uses a large basis of
interpolators $O_i,\, i = 1,2, \, ... \, N$ and considers the generalized
eigenvalue problem of the $N \times N$ correlation matrix $C(t)$,
\begin{equation}
C(t) \, \vec{v}^{\ (n)} \; = \; \lambda(t)^{(n)}\ C(t_0)\ \vec{v}^{\ (n)} 
\; . 
\label{genevalprob}
\end{equation}
The ordered eigenvalues, 
$\lambda^{(1)} > \lambda^{(2)} > \lambda^{(3)} > \, ... \, $, behave as
\cite{variation2}
\begin{equation}
\lambda^{(n)}(t) \; = \; e^{-(t-t_0)W_n}\ [1+\mathcal{O}(e^{-(t-t_0)\Delta_n})] \; ,
\label{evals}
\end{equation}
where $W_n$ is the energy of the $n$-th state and $\Delta_n$ the distance 
of $W_n$ to the neighboring energy level. Effectively the ground and excited
states are disentangled, and each energy appears in an individual eigenvalue.
The variational method can only be as good as the basis of interpolators 
$O_i$ one uses and different construction principles were proposed
\cite{hadrons}.  

Even with the variational method, the calculation of excited states is a
challenging enterprise. Several non-trivial steps are involved: One has  
to construct a suitable basis of interpolators and select from those a subset
which optimizes the quality of the signal, i.e., provides the cleanest
effective mass plateaus. For an extrapolation to the infinite volume,
continuum and chiral limits, the signals from several lattices with different
size, coupling and mass have to be matched. Here it turns out that the
eigenvectors $\vec{v}^{\ (n)}$ provide an important marker for identifying
the states. Finally, in a fully dynamical simulation hadrons may decay. This
leads to coexistence of bound- and scattering states which have similar 
energies $W_n$ and thus mix in the spectral sums (\ref{twopoint}). Because
scattering states have relative momentum, they can be identified by a finite
volume analysis, since the minimal possible momentum is inverse proportional
to the spatial extent of the lattice \cite{scattertechniques}. 

Although the methods for solving the technical problems of an excited
spectroscopy calculation are known, so far the typical analysis produced only
the masses for one or two excitations and most papers are still in the
quenched approximation, where hadrons cannot decay.

There is a long standing tradition of studying problems of theoretical physics
in low dimensional quantum field theories. This allows to test new concepts
and techniques in an environment where often the situation is simpler and,
when the lattice is used, large statistics is easy to generate (examples for
such 2-d studies are \cite{variation2}, \cite{more2d}). The experience gained
in 2-d is then often an important guideline for the more demanding case of
full QCD.

In this paper we follow this tradition and present first results of a lattice 
study of excited states in the 2-d Gross-Neveu model \cite{grossneveu}. This
model is particularly attractive for excited state spectroscopy, since in the
infinite flavor limit, analytic results suggest that the masses of the
excitations are integer multiples of the ground state mass. Our project aims
at extracting with the variational method as many excitations as possible and
to study their volume dependence to learn about scattering. 

\section{Lattice simulation of the Gross-Neveu model}

\subsection{Setting of the calculation}
\noindent
We use the Wilson action for discretizing the fermions on a 2-d 
lattice $\Lambda$. The total action is
\begin{equation}
S[\ov{\psi},\psi, \Phi] \; = \; 
\sum_{\vec{x},\vec{y} \in \Lambda} \ 
\ov{\psi}(\vec{x}) \ D(\vec{x},\vec{y}) \ \psi(\vec{y}) \; + \; 
\frac{1}{2} \ \sum_{\vec{x}\in \Lambda} \ \Phi(\vec{x})^2 \ .
\end{equation}
The fields $\ov{\psi},\psi$ are vectors of $N_f$ two-component spinors, one
for each flavor, and we use matrix/vector notation for all indices. $\Phi$
denotes a one-component real scalar field, which generates the 4-fermi interaction of the
model through a Hubbard-Stratonovich
transformation. The Dirac matrix
$D$ is diagonal in flavor space and reads
\begin{equation}
D(\vec{x},\vec{y}) \; = \;  [\ m + \sqrt{g}\ \Phi({n})\ ] \,
\delta_{\,\vec{x},\vec{y}} \;  - \, \sum_{\mu=\pm 1}^{\pm 2}
\frac{\mathds{1} \mp \gamma_\mu}{2} \delta_{\,\vec{x}+\hat{\mu},\vec{y}} \; .
\end{equation}
The matrices $\gamma_\mu$ are given by the Pauli matrices, $\gamma_1 = \sigma_1,
\gamma_2 = \sigma_2, \gamma_5 = \sigma_3$. 

The dynamical  
simulation of the model is done with standard Hybrid Monte Carlo methods
and we work with $N_f = 2,4$ and 6 flavors. Our $L_1 \times L_2$ lattices have
sizes $L_1 = 10 \, ... \, 24$, $L_2 = 32 \, ... \, 64$ and we typically use
four values of the coupling
constant $g$ and several different masses $m$. Statistical errors are
determined with the Jackknife-method. We find that statistics of a few hundred
configurations per ensemble are sufficient for statistical 
errors in the one percent range. Throughout this paper we set the 
lattice constant equal to one and all results are in lattice units.

\subsection{Interpolators}
\noindent
For our analysis of excited states we use the following 
set of flavor singlet interpolators:
\begin{eqnarray}
O_1(t) &=& \overline{\psi}(x,t)\ \gamma_5 \ \psi(x,t) \; ,  
\label{interpols}
\\
O_2(t) &=& \overline{\psi}(x,t)\ \gamma_1 \ \psi(x,t) \; , 
\nonumber \\
O_3(t) &=& \frac{1}{2}\ \overline{\psi}(x,t)\ 
\gamma_5 \ [\psi(x+n,t)+\psi(x-n,t)] \; , 
\nonumber \\
O_4(t) &=& \frac{1}{2}\ \overline{\psi}(x,t)\ 
\gamma_1 \ [\psi(x+n,t)+\psi(x-n,t)] \; ,
\nonumber \\
O_5(t) &=& \frac{1}{4}\ [\overline{\psi}(x+m,t) 
+ \overline{\psi}(x-m,t)]\ \gamma_5 \  [\psi(x+n,t)+\psi(x-n,t)] \; ,
\nonumber \\
O_6(t) &=& \frac{1}{4}\ [\overline{\psi}(x+m,t) 
+ \overline{\psi}(x-m,t)]\ \gamma_1 \  [\psi(x+n,t)+\psi(x-n,t)] \; ,
\nonumber \\
O_8(t) &=& \frac{1}{4}\ [\overline{\psi}(x+m,t) 
+ \overline{\psi}(x-m,t)]\ [\psi(x+n,t)-\psi(x-n,t)] \; ,
\nonumber \\
O_9(t) &=& \frac{1}{4}\ [\overline{\psi}(x+m,t) 
- \overline{\psi}(x-m,t)]\ \gamma_5 \ [\psi(x+n,t)-\psi(x-n,t)] \; ,
\nonumber \\
O_{10}(t) &=& \frac{1}{4}\ [\overline{\psi}(x+m,t) 
- \overline{\psi}(x-m,t)]\ \gamma_1 \ [\psi(x+n,t)-\psi(x-n,t)] \; .
\nonumber
\end{eqnarray}
To project to vanishing total momentum, we sum over the spatial index $x$, a
step which we suppress in the list of Eq.~(\ref{interpols}) for notational
convenience. The interpolators are built from field variables at different
lattice sites, and the integer valued parameters $n$ and $m$ determine how
many steps in spatial direction the field variables are displaced.

Furthermore it is possible to have a relative minus sign between the 
displaced fields which gives rise to
a derivative type of fermion source. In order to have the same quantum numbers
as the combination with a plus sign, an additional factor of $\gamma_1$
appears. In 2-d all products of $\gamma$-matrices can be simplified to a
single $\gamma_\mu$, or $\mathds{1}$ due to the algebra of the Pauli matrices. 

Our interpolators have negative parity, i.e., they acquire a minus sign under
the parity transformation
\begin{equation}
\psi(x,t) \stackrel{\mf{P}}{\longrightarrow} 
\psi(x,t)^{\mf{P}}=\gamma_2\ \psi(-x,t) \; \; , \; \; 
\overline{\psi}(x,t) \stackrel{\mf{P}}{\longrightarrow}
\overline{\psi}(x,t)^{\mf{P}}=\overline{\psi}(-x,t)\ \gamma_2 \ .
\end{equation}
Furthermore all interpolators are eigenstates with $\mf{C} = -1$
of the charge conjugation 
\begin{equation}
\psi(x,t) \stackrel{\mf{C}}{\longrightarrow} 
\psi(x,t)^{\mf{C}}=C^{-1}\ \overline{\psi}(x,t)^T \; \; , \; \; 
\overline{\psi}(x,t) \stackrel{\mf{C}}{\longrightarrow} 
\overline{\psi}(x,t)^{\mf{C}}=-\psi(x,t)^T\ C \; ,
\end{equation}
where $C$ is the charge conjugation matrix obeying 
$C\gamma_{\mu}C^{-1}=-\gamma_{\mu}^T$, which can be chosen as  $C=i\gamma_2$.
We stress that in 2-d no angular momentum exists and thus $P$ and $C$
completely characterize our flavor singlet states.

In our analysis we use two different choices for the displacement parameters
$n$ and $m$. We set $n = 3$ and $m = 3$ to obtain the 9 
interpolators $O_1$ ... $O_{10}$ as listed in (\ref{interpols}) 
(note that there is no interpolator $O_7$). A second choice, 
$n = 4$ and $m = 2$, gives rise to our interpolators $O_{11}$ ... $O_{20}$
labeled accordingly. Thus we can use correlation matrices with a maximum size
of $16 \times 16$.

The implementation of the correlation matrix was checked in two independent
programs and compared to the results for the free case computed from Fourier
transformation. Within error bars, the correlation matrix was found to be
hermitian as expected. The decay properties of the individual entries in the
correlation matrix are either of the $\cosh$- or $\sinh$-type.

\section{Results}

\noindent
For a first assessment of the data we consider plots of the correlators and of
effective energies, a term which is more suitable than the usual 
``effective masses'', since some of our states will turn out to be scattering
states, such that their energy does not correspond to a single mass. 
The effective energies are defined as
\begin{equation}
\label{effmass}
W^{\text{eff}}(t+1/2) \; = \; \ln \frac{c(t)}{c(t+1)} \; ,
\end{equation}
where $c(t)$ is either an individual entry of the correlation matrix, or one
of the eigenvalues of the generalized eigenvalue problem (\ref{genevalprob}).

\begin{figure}[t]
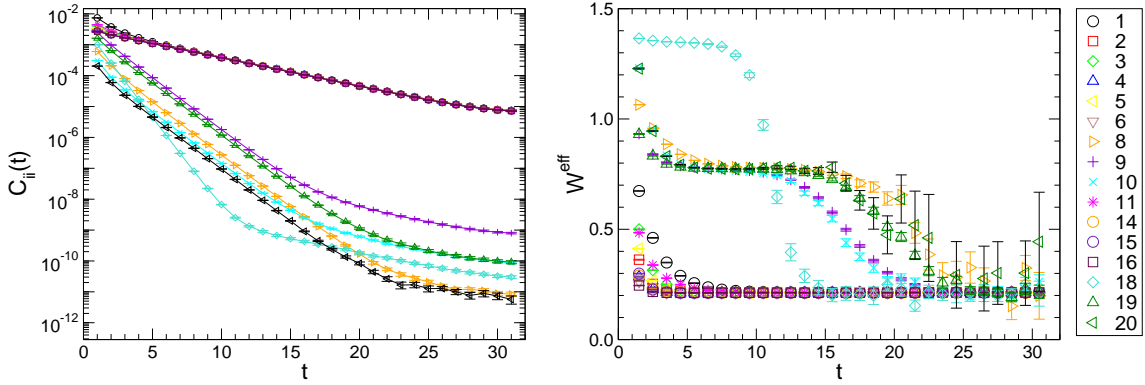

\begin{center}
  \includegraphics*[scale=0.34]{diagonal__11111101110011110111.eps} 
\hfill
  \includegraphics*[scale=0.34]{singlecorr_11111101110011110111.eps}
\end{center}
\caption{\label{fig:corr} The diagonal entries $C_{ii}(t)$ of the correlation
  matrix (l.h.s.~plot) and the corresponding effective energies (r.h.s) 
for a $16\times64$ lattice with $g=0.05$, $m=0.05$ and $N_f=2$. The numbers in
  the legend are the labels of our interpolators.}
\end{figure}

In the l.h.s.~plot of Fig.~\ref{fig:corr} we show the diagonal entries 
$C_{ii}(t)$ of the correlation matrix on a logarithmic scale. The r.h.s.~plot
contains the corresponding effective energies. The data are from $16\times64$ 
lattices at $g=0.05$, $m=0.05$ and $N_f=2$.

It is obvious from the l.h.s.~plot that on the logarithmic scale 
the correlators $C_{ii}(t)$ fall on a
straight line beyond $t \sim 3$, i.e., are dominated by a single
exponential. It is, however, very remarkable, that up to $t \sim 20$ they have
rather different slopes, and only beyond that value they all settle for the
smallest slope. In particular the interpolators with a relative minus sign 
(derivative interpolators) show a steeper slope compared to the 
interpolators with relative plus signs, indicating that the former strongly 
couple to excitations. We stress, however, that also the derivative 
interpolators couple to the ground state, as can be seen in the 
effective energy plots (r.h.s.): They clearly show a second plateau at 
the ground state energy for large values of $t$. Interpolator $O_{18}$ 
even seems to couple to a higher excited state, but settles at the 
ground state energy beyond $t\approx 13$. The fact, that some of 
our interpolators couple very strongly to excitations shows that they 
have a large overlap with the true physical states. It must, however, be
understood, that the situation in 2-d is particularly simple due to the lack
of angular momentum. All a spatial wave function can do is to have nodes in
the single spatial direction. Such wave functions are easily obtained by 
linear combinations of our interpolators.

\begin{figure}[t]
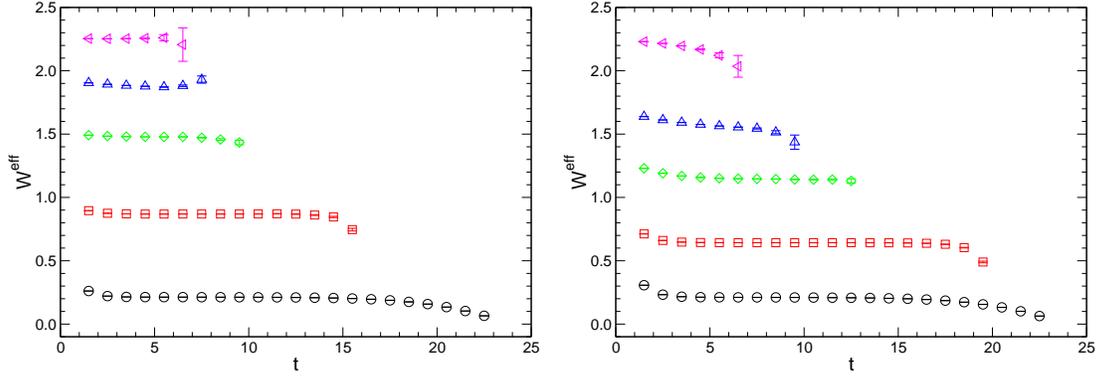

\begin{center}
  \includegraphics*[scale=0.29]{masses_00101001000000000110_14.eps} 
\hspace{1mm}
  \includegraphics*[scale=0.29]{masses_00101001000000000110_20.eps}
\end{center} 
\caption{\label{fig:eff} 
Effective energies from the eigenvalues of a $5\times5$
correlation matrix. We compare results from a $14\times48$ lattice (l.h.s.)
to the data from $20\times48$ (r.h.s.), both for $g=0.05$, $m=0.05$ 
and $N_f=2$.}
\end{figure}

Let us now come to the analysis of the results from the variational analysis.
In Fig.~\ref{fig:eff} we show the effective energies for the eigenvalues from
the generalized eigenvalue problem for a $5 \times 5$ correlation matrix built
from the interpolators $O_3, O_5, O_8, O_{18}$ and $O_{19}$ for 
$g=0.05$, $m=0.05$, $N_f=2$. We compare two different volumes, 
$14\times48$ in the l.h.s.~plot and $20\times48$ on the r.h.s.

The plots show long reliable plateaus, in particular for the three lowest
states. The two highest states on the larger lattice (r.h.s.) 
show noticeable deviations from a horizontal line, which might be an effect
of mixing with other states through the correction term in (\ref{evals}).
 
It is interesting to note that except for the lowest lying state (circles) all
states show a pronounced volume dependence and are shifted towards smaller
energies for the larger volume. This indicates that the higher states might be
scattering states, which show, as discussed, a pronounced volume
dependence. In a future contribution a more detailed analysis of the volume
effects will be presented, which establishes that indeed the excited states
of Fig.~\ref{fig:eff} are scattering states.

Here we would like to focus on a different aspect: For the finite volume
analysis it is important to compare the energies from several different
lattice volumes. However, when changing the volume, the spectrum is shifted
and it is a non-trivial task to identify the individual states, in particular
since two states might have changed their relative position. For the
purpose of matching states on different volumes, the eigenvectors of the
generalized eigenvalue problem (\ref{genevalprob}) are an important tool. 
Since they contain the information on how a state is composed from the
individual interpolators, they serve as a ``fingerprint'' 
for an individual state. 

\begin{figure}[t]
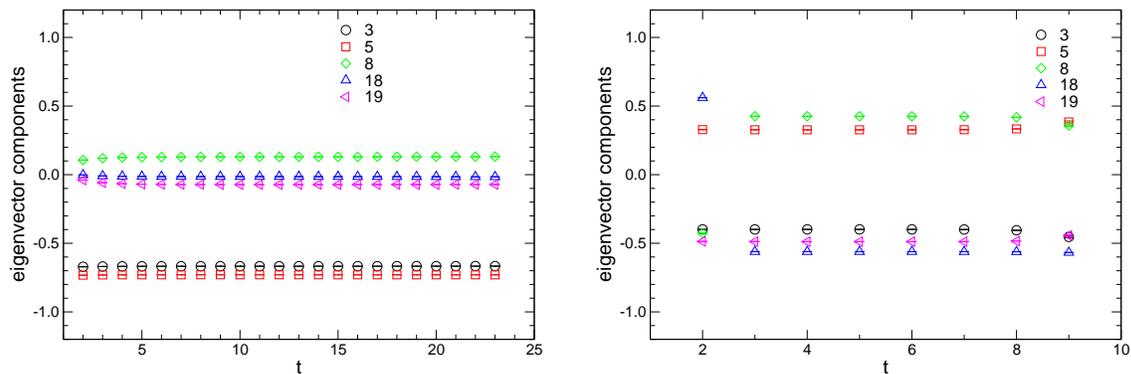

\begin{center}
  \includegraphics*[scale=0.29]{vectors_00101001000000000110_1.eps} 
\hspace*{0.5cm}
  \includegraphics*[scale=0.29]{vectors_00101001000000000110_5.eps}
\end{center} 
\caption{\label{fig:ei} Entries of the eigenvectors for the ground state
  (l.h.s) and the fourth excitation (r.h.s) as a function of $t$ 
($16\times48$, $g=0.01$, $m=0.05$, $N_f=2$). The numbers in
  the legend are labels of the interpolators.}
\end{figure}

In Fig.~\ref{fig:ei} we display the entries of the
eigenvectors (normalized to 1)
for the ground state (l.h.s.) and our highest excited state 
(r.h.s.) as a function of $t$ ($16 \times 48$, $g=0.01$, $m=0.05$, $N_f=2$).
Also the eigenvector entries form pronounced plateaus and the position of the
plateau gives the coefficient for the corresponding interpolator in the linear
combination which builds up the states. These coefficients are surprisingly
stable for different volumes and thus are suitable quantities to be used in
the identification of the states. Furthermore, the coefficients provide
physically relevant information on how the states are composed: 
The ground state (l.h.s.) is dominated by the operators $O_3$ and $O_{5}$, 
whereas the highest excited state (r.h.s.) receives important contributions
also from the derivative interpolators. Such an analysis allows for a
qualitative understanding of the nature of the excited states.
In an upcoming paper we will use the methods presented here for a detailed
analysis of the spectrum of excitations in the lattice Gross Neveu model.

\begin{thebibliography}{99}

\bibitem{hadrons}
  C.~McNeile,
  PoS(LATTICE 2007)019 [arXiv:0710.0985 [hep-lat]];
%
  T.~Burch {\it et al},
  Phys.\ Rev.\  D {\bf 74} (2006) 014504;
  Phys.\ Rev.\  D {\bf 73} (2006) 094505;
  Phys.\ Rev.\  D {\bf 73} (2006) 017502;
  Phys.\ Rev.\  D {\bf 70} (2004) 054502;
%
  S.~Basak {\it et al.},  arXiv:0709.0008 [hep-lat];
  Phys.\ Rev.\  D {\bf 72} (2005) 074501;
  Phys.\ Rev.\  D {\bf 72} (2005) 094506;
%
  D.~Br\"ommel {\sl et al}
  Phys.\ Rev.\  D {\bf 69} (2004) 094513;
%
  S.\ Sasaki, T.\ Blum and S.\ Ohta, Phys.\ Rev.\ D65 (2002) 074503;
%
  K.~Sasaki, S.~Sasaki and T.~Hatsuda,
  Phys.\ Lett.\ B {\bf 623} (2005) 208;
%
  J.~M.~Zanotti {\it et al.},
  Phys.\ Rev.\ D {\bf 68} (2003) 054506;
%
  W.~Melnitchouk {\it et al.},
  Phys.\ Rev.\ D {\bf 67} (2003) 114506;
%
  N.~Mathur {\it et al.},
  Phys.\ Lett.\ B {\bf 605} (2005) 137;
%
  F.~X.\ Lee {\it et al.},
  Nucl.\ Phys.\ Proc.\ Suppl.\ {\bf 119} (2003) 296;
%
  S.~Sasaki,
  Prog.\ Theor.\ Phys.\ Suppl.\  {\bf 151} (2003) 143.


\bibitem{otherextract}
  G.~Fleming PoS(LATTICE2007)096;
%
  G.M.~von Hippel, R.~Lewis and R.G.~Petry,
  PoS(LATTICE 2007)043 [arXiv:0710.0014 [hep-lat]];
%
  D.~Guadagnoli, M.~Papinutto and S.~Simula,
  Phys.\ Lett.\  B {\bf 604} (2004) 74;
%
  G.P.\ Lepage, Nucl.\ Phys.\ (Proc.\ Suppl.) {\bf 106} (2002) 12;
%
  M.\ Asakawa, T.\ Hatsuda and Y.\ Nakahara,
  Prog.\ Part.\ Nucl.\ Phys.\ {\bf 46} (2001) 459.

\bibitem{variation1}
  C.~Michael, Nucl.~Phys.~\textbf{B259} (1985) 58.

\bibitem{variation2}
  M.~L\"uscher and U.~Wolff, Nucl.~Phys.~\textbf{B339} (1990) 222.

\bibitem{scattertechniques}
  M.~L\"uscher,
  Nucl.\ Phys.\  B {\bf 364} (1991) 237;
%
  Nucl.\ Phys.\  B {\bf 354} (1991) 531.


\bibitem{more2d} 

  W.~Bietenholz, S.~Shcheredin and J.~Volkholz,
  arXiv:0710.0997 [hep-lat];
%
  U.~Wolff,
  arXiv:0707.2872 [hep-lat];
%
  C.~Gattringer, V.~Hermann and M.~Limmer,
  Phys.\ Rev.\  D {\bf 76} (2007) 014503;
%
  C.~Gattringer and I.~Hip,
  Phys.\ Lett.\  B {\bf 480} (2000) 112;
%
  C.B.~Lang and T.K.~Pany,
  Nucl.\ Phys.\  B {\bf 513} (1998) 645;
%
  C.~Gattringer, I.~Hip and C.B.~Lang,
  Phys.\ Lett.\  B {\bf 409} (1997) 371;
%
  C.~Gattringer and C.B.~Lang,
  Nucl.\ Phys.\  B {\bf 391} (1993) 463;
%
  W.~Bietenholz, F.~Hofheinz and J.~Nishimura,
  JHEP {\bf 0209} (2002) 009;
%
  W.~Bietenholz, A.~Pochinsky and U.J.~Wiese,
  Phys.\ Rev.\ Lett.\  {\bf 75} (1995) 4524.


\bibitem{grossneveu}
D.J.~Gross and A.~Neveu, Phys.~Rev.~\textbf{D10} (1974) 3235.


\end{thebibliography}
\end{document}